\documentclass[preprint]{elsarticle}
\usepackage[english]{babel}
\usepackage{amssymb, amsmath}
\usepackage{booktabs}
\usepackage{graphicx}
\usepackage{url}
\pdfoutput=1        
\usepackage{float} 
\usepackage{pifont} 
\usepackage{natbib} 
\usepackage{geometry} 
\usepackage{graphicx} 
\usepackage{hyperref} 

\begin{document}

\begin{frontmatter}
\title{Optimal strategies for observation of active galactic nuclei variability with Imaging Atmospheric Cherenkov Telescopes}
\author[desy]{Matteo Giomi}
\ead{matteo.giomi@desy.de}
\author[desy]{Lucie Gerard}
\author[desy]{Gernot Maier}
\address[desy]{Deutsches Elektronen-Synchrotron DESY, D-15738 Zeuthen, Germany}
\begin{abstract}
Variable emission is one of the defining characteristic of active galactic nuclei (AGN). While providing precious information on the nature and physics of the sources, variability is often challenging to observe with time- and field-of-view-limited astronomical observatories such as Imaging Atmospheric Cherenkov Telescopes (IACTs). In this work, we address two questions relevant for the observation of sources characterized by AGN-like variability: what is the most time-efficient way to detect such sources, and what is the observational bias that can be introduced by the choice of the observing strategy when conducting blind surveys of the sky. Different observing strategies are evaluated using simulated light curves and realistic instrument response functions of the Cherenkov Telescope Array (CTA), a future gamma-ray observatory. We show that strategies that makes use of very small observing windows, spread over large periods of time, allows for a faster detection of the source, and are less influenced by the variability properties of the sources, as compared to strategies that concentrate the observing time in a small number of large observing windows. Although derived using CTA as an example, our conclusions are conceptually valid for any IACTs facility, and in general, to all observatories with small field of view and limited duty cycle.
\end{abstract}

\begin{keyword}
Cherenkov Telescopes \sep Active Galactic Nuclei \sep variability \sep gamma-ray \sep observation planning \sep extragalactic sky survey
\end{keyword}

\end{frontmatter}

\section{Introduction}

Transient and variable emission has been observed from gamma-ray sources such as binary systems, active galactic nuclei and gamma-ray bursts. Variability studies shed light on physical processes responsible for the acceleration of charged particles and the photon emission in these objects. The determination of the characteristic physical time scales for particle acceleration and gamma-ray emission mechanisms, and their dependence on the relativistic beaming, allow us to gain an understanding of the origin of flares, and the size, structure, and location of the emission region. 

The majority of variable gamma-ray sources detected at very-high-energy (VHE, from $\sim50$~GeV to hundreds of TeV) are active galactic nuclei (AGN), powered by supermassive black holes accreting matter at the center of some galaxies. Most of the gamma-ray bright AGN are blazars, a subclass of AGN characterized by the presence of two collimated outflows of relativistic plasma (jets), streaming away from the central black hole along the line of sight, so that one jet is pointed towards the Earth~\cite{urry_padovani}. Relativistic beaming of the emission of the jet enhances the luminosity of the source and shifts it to higher energies, making these sources bright VHE emitters; in fact, blazars account for $\sim1/3$ of all the sources currently detected at these energies. Blazar emission spans through the entire electromagnetic spectrum and it is characterized by a strong temporal variability, observed on time scales ranging from years down to minutes.

The observation of variable emission patterns is challenging with IACTs, due to their small field of view and the fact that observations are limited to dark nights and good weather periods. For these reasons, observing strategies for IACTs must be carefully optimized in order to achieve their goal in as little time as possible. The common approach to observation planning makes use of the sensitivity to describe the performance of the instrument and compare it to the mean or low-state flux of the sources. This approach is not optimized to plan observations of strongly variable sources as it does not explicitly take into account the variability. In this work, variability is taken into account by the simulation of light curves similar to those measured for AGN. Various types of variability properties are considered. Different observing strategies are modeled and, for each one, the probability of detecting the sources is computed. This probability provides a measure of the performance of the observing strategy and is used to quantitatively compare different strategies. Two questions that are relevant in the context of observation planning are addressed: what is the fastest way to detect a weak, variable source with given variability characteristics, and which observing strategies are more suited to conduct variability-unbiased, blind-sky surveys.

Our work is applied here to optimize AGN observations with the Cherenkov Telescope Array (CTA), a future VHE gamma-ray observatory ~\cite{cta_design}. CTA will consist of 80-100 of telescopes of three different sizes, resulting in a ten times better sensitivity than any of the current IACT arrays and a wide, four-orders-of-magnitude, energy range. The observation of AGN will be one of the major goals for CTA, and simulations show that CTA will detect a large number ($\sim10^3$) of these objects~\cite{cta_agn}.

This paper is structured as follows: Section~\ref{sec:lc_simu} presents the simulation of AGN variability. In Section~\ref{sec:obs_strat_model}, the model of the observing strategies is illustrated. Section~\ref{sec:obs_quant} describes the method used to compute the performance of the observing strategies for a given set of variability properties. Finally, in Section~\ref{sec:results} the optimal strategies to observe known sources and to conduct a blind-sky survey are identified.

\section{Simulation of VHE AGN variability}\label{sec:lc_simu}

The variability of a source is usually presented through its light curve, showing the flux in a given energy band as a function of time. VHE AGN light curves appear aperiodic. Flux variations down to the timescale of minutes have been observed for the brightest sources such as Mkn 421~\cite{mkn_421_var}, PKS 2155-304~\cite{pks_2155304_var}, and Mkn 501~\cite{mkn_501_magic}.

As of today, little information is available on the temporal properties of VHE AGN light curves. To attempt a description of the variability characteristics of the whole class of AGN, data collected at other wavelengths have to be considered. In particular we will refer to results from the \textit{Fermi} Large Area Telescope (LAT)~\cite{fermi_sat}, observing the sky in the high-energy (HE, from tens of MeV to hundreds of GeV) band. By doing this we assume that variability properties observed in the HE band are representative of those in VHE. This working assumption might be supported since similar physical phenomena could be responsible for the emission in these two adjacent energy bands. However, it is worth stressing that the light curves we simulate are meant only to give a good representation of the available data; no attempt is made here to a detailed model of VHE AGN variability.

Fourier analysis is one of the most common and powerful tools used to characterize time series. Referring to its properties in the frequency domain, AGN variability is often described as power-law noise, i.e. the Power Spectral Density (PSD) of AGN light curves is well reproduced by a power-law function of the frequency $\nu$, $P(\nu)\propto\nu^{-\beta}$, over a wide range of frequencies. Fourier spectral analysis has been performed on a sample of 156 BL Lacs and 56 Flat Spectrum Radio Quasars (FSRQs) in the second \textit{Fermi} LAT AGN catalog (2LAC)~\cite{fermi_2lat_cat}. In this case monthly binned light curves of the integral flux above 100 MeV are used, finding $\beta\sim1.15\pm0.1$ for both source classes in the frequency range $[0.033,~0.5]~\mbox{month}^{-1}$. Similar analysis performed on a smaller sample of bright sources (22 FSRQs and 6 BL Lacs) for which three- and four-days binned light curves could be produced, lead to values of $\beta$ of 1.7 and 1.5 for BL Lacs and FSRQs respectively~\cite{fermi_lbas_var}. In the VHE energy domain the only case in which Fourier analysis of the light curve has been performed is the PKS 2155-304 2006 flare observed by H.E.S.S. The PSD for the 1 minute binned light curve is compatible at $90\%$ confidence level with a power law of index $\beta=2$ in the frequency range $\sim[10^{-4},10^{-3}]$ Hz~\cite{pks_2155304_var}.

Another important feature of AGN light curves is a linear relation between the root mean square (RMS) amplitude and the mean flux of groups of contiguous bins of the light curve. This proportionality implies that fluctuations around the mean value, i.e. variability, are enhanced when the flux is higher. This has been observed in X-ray for many AGN and other accreting objects, like binary systems~\cite{vaughan, cygnus3var}. At VHE such behavior has been detected in the well studied PKS 2155-304~\cite{pks_2155304_var_all} and Mkn 421~\cite{combined_lcs}. As observed by Uttley and McHardy~\cite{uttley_nonlinear}, and Biteau and Giebels~\cite{biteau}, this RMS-flux proportionality represents a strong constraint on the models that must be used to reproduce AGN light curves. In particular, these authors have demonstrated that only models in which the simulated flux is obtained as the exponential of an underlying stochastic process are able to produce such a relation.

PSD properties and the RMS-flux relation define the variability profile of the light curves; the variance of the light curve sets the scale of the fluctuations. This quantity is more commonly expressed through the use of the fractional RMS amplitude $F_{\text{rms}}$\footnote{$F_{\text{rms}}=\sqrt{\sigma_{\phi}^2}\slash\langle\phi\rangle$, where $\langle\phi\rangle$ and $\sigma_{\phi}^2$ are respectively the mean flux and variance of the light curve. This quantity expresses, independently from the source strength, the amount of variability in the light curve. When dealing with real data, the variance  $\sigma_{\phi}^2$ is more conveniently replaced by the excess variance~\cite{vaughan}.}. Fractional RMS amplitudes ranging from 20\% to 60\% have been observed in the VHE light curve above 200 GeV of PKS 2155-304~\cite{pks_2155304_var_all}. Values of $F_{\text{rms}}$ for most of the \textit{Fermi} 2LAC sources are found to vary between 20\% and 80\%~\cite{fermi_2lat_cat}.

We simulate AGN-like light curves of the flux above 100 GeV, the typical energy threshold for VHE observatories. To produce time series exhibiting a power-law PSD and a linear RMS-flux relation, we follow the method proposed by Uttley and McHardy~\cite{uttley_nonlinear}. The light curve $\phi(t_i)$ is obtained through an exponential transformation of a linear, aperiodic time series with zero mean, $x(t_i)$:
\begin{equation}\label{eqn:exp_lc}
 \phi(t_i)=\exp{[x(t_i)]} 
\end{equation}
The Timmer and K\"{o}nig algorithm~\cite{timmerkoenig} is used to produce the input time series $x(t_i)$. This algorithm produces linear time series with a power-law PSD of arbitrary slope. It must be noted that, as a consequence of the exponential transformation, the PSD of the resulting light curve $\phi(t_i)$ is different from that of the input time series $x(t_i)$. However, for the broad continuous PSD observed in AGN, this distorting effect is relatively small and can be neglected~\cite{uttley_nonlinear}.

The exponential transformation of Eq.~\ref{eqn:exp_lc} enhances points above the mean while reducing points below the mean. As a consequence, when the variance of the Timmer and K\"{o}nig time series $x(t_i)$ increases, flares in the light curve $\phi(t_i)$ becomes more prominent compared to the dips. The aspect of the resulting light curves $\phi(t_i)$ is thus influenced by the variance of the input time series $x(t_i)$, a parameter that is inherent to the light curve simulation algorithm but has no clear physical interpretation\footnote{A possible workaround to this problem would be to assume a log-normal distribution for VHE AGN fluxes. Under this assumption, the statistical properties of the light curves can be related to the mean and variance of the input series $x(t_i)$, eliminating the need of this additional parameter. However, as of today, log-normal distributions of AGN fluxes have not been observed in VHE, if the entire light curve of the source is considered (see, for example, the bi-modal distribution of PKS 2155-304 flux over the 2005-2007 period as observed by H.E.S.S.~\cite{pks_2155304_var_all}). Although log-normality at the highest energies is not ruled out, due to the generally poor sampling flux states of the sources, biased towards the observation of flares, we still prefer not to use this additional hypothesis.}. In this work this parameter is fixed for all the light curves: the Timmer and K\"{o}nig series are normalized to $F_{\text{rms}}=50\%$ before taking the exponential transformation. Different normalization for the Timmer and K\"{o}nig series have been tested and shown negligible impact on the conclusion of this work.

The choice of the frequency, i.e. the spacing between two contiguous points, and length, of the simulated light curves defines the range of Fourier frequencies that contributes to the signal. These parameters are chosen in such a way as to include contributions from all the frequency components that are currently probed by HE and VHE observations. The lowest frequency considered is $0.033$ month$^{-1}$, probed by \textit{Fermi}-LAT for sources in the second LAT AGN catalog~\cite{fermi_2lat_cat}. The highest frequency will be 1 minute$^{-1}$, as in the case of PKS 2155-304 2006 flare light curve measured by H.E.S.S. To cover this frequency range, we generate 30-months-long light curves with one point per minute. From these long light curves we extract 100-days-long segments at random positions. In this way the light curves will have a duration that is comparable to the yearly visibility of a source (see Section~\ref{subsec:visibility} for details), while still including variability from all the frequency components that have been observed. This choice does not completely eliminate the problem of \lq red noise leak\rq~(see ~\cite{vaughan} and references therein), as AGN light curves contain variability on time scales longer than 30 months. To mitigate this effect one can generate much longer light curves, extrapolating the PSD to very low frequencies. Since low frequency behavior of the PSD of VHE AGN light curves is not known (if $\beta \geq 1$ a flattening must be present at low frequencies to avoid divergence of the power in the signal), and considering that the effect of red noise leak is negligible for $\beta \leq 1.5$~\cite{vaughan}, in this work it has been chosen not to correct the light curves for this effect.

\begin{figure}[htbp]
	\begin{minipage}[b]{\textwidth}
	\centering
	\includegraphics[height=0.28\textheight]{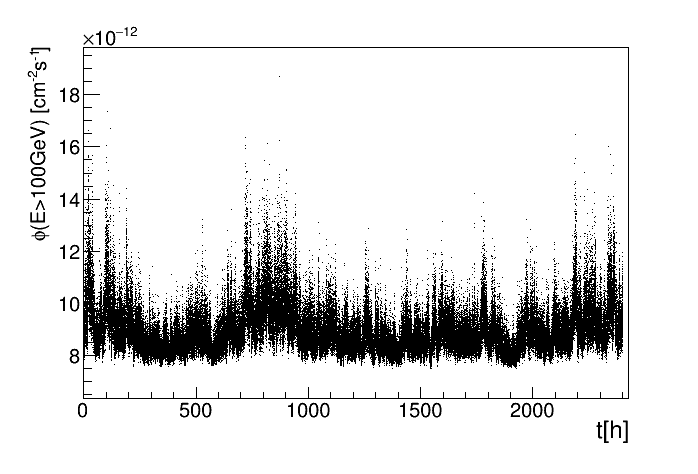}
	\end{minipage}
	\\
	\\
	\begin{minipage}[b]{\textwidth}
	\centering
	\includegraphics[height=0.28\textheight]{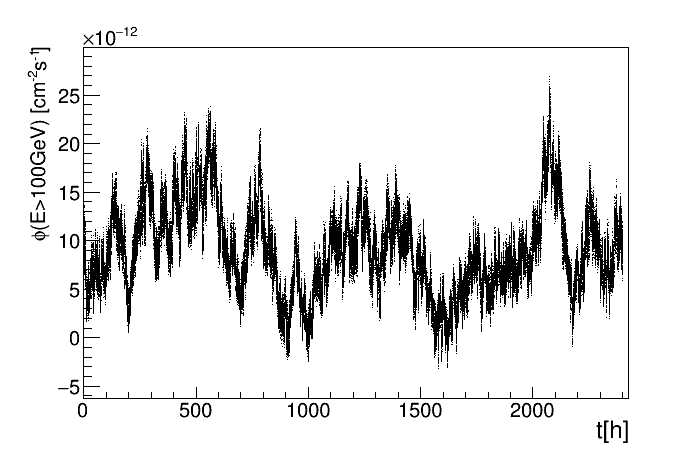}
	\end{minipage}
	\\
	\\
	\begin{minipage}[b]{\textwidth} 
	\centering
	\includegraphics[height=0.28\textheight]{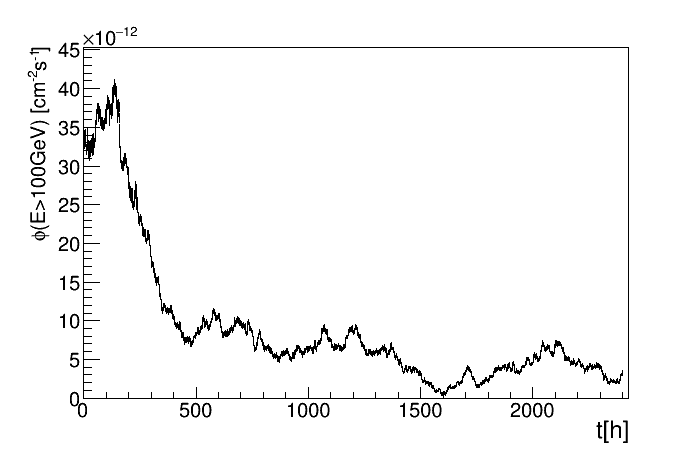}
	\end{minipage}
	\caption{Examples of the light curves used in this work: \textit{top}: low-variability light curve ($\beta=1.0$, $F_{\text{rms}}=10\%$), \textit{middle}: medium-variability light curve ($\beta=1.5$, $F_{\text{rms}}=50\%$), \textit{bottom}: highly-variable light curve ($\beta=2.0$, $F_{\text{rms}}=100\%$).}
	\label{fig:lc_examples}
\end{figure}

To simulate different kinds of variability we consider variations of $\beta$ and $F_{\text{rms}}$, assuming a flat distribution of sources with respect to both of these parameters. We vary the PSD index $\beta$ between 1 and 2 with an increment of $0.1$. $F_{\text{rms}}$ will take values between 10\% and 100\% using steps of 10\%. For each combination of $\beta$ and $F_{\text{rms}}$, a set of 1000 light curves is produced. All light curves in each set share the same variability characteristics. The mean flux of all the light curves is set to 1\% of the flux above 100 GeV of the Crab Nebula\footnote{$\phi_{\text{Crab}}(E>100\mbox{GeV})=9\cdot 10^{-10}\mbox{cm}^{-2}\mbox{s}^{-1}$, as computed from the pure power-law parametrization of the spectrum of the Crab Nebula given in~\cite{hess_crab}}, representing the target AGN population for CTA. Figure~\ref{fig:lc_examples} shows some examples of the light curves thus produced, illustrating the different kinds of variability considered in this work.

\section{Model of the observing strategies}\label{sec:obs_strat_model}

A model of the observation of a source has been set up. This model includes source visibility, the pointing scheme used to carry out the observations, and the characteristic of the instrument.

\subsection{Source visibility}\label{subsec:visibility}

The visibility of the source is taken into account, using visibility parameters that matches realistic observing conditions of IACTs. The visibility windows are defined as the time intervals in which the sun is 15 degrees below the horizon, the moon phase is less than 30\%, and the altitude of the source is greater than 60 degrees. They are computed using the \texttt{astropy}~\cite{astropy}, and \texttt{PyEphem}~\cite{pyephem} python packages.
To simplify the modeling of the pointing schemes, and to be able to test observing strategies that makes use of large observing windows, we discard visibility windows that are smaller than 4 hours. The resulting visibility profile spans 100 days, with 4 blocks of 8 to 12 consecutive days where the source can be observed for $\sim4.5$ hours, separated by periods of bright moon. It is computed for a source whose declination ($\delta$) is equal to the latitude ($\phi$) of the observatory, but it is a good approximation for all the sources for which $|\delta-\phi|\leq 10^{\circ}$. In the following, we will refer to the visibility windows simply as nights.

\subsection{Pointing schemes}

A pointing scheme specifies the series of time intervals in which the instrument is observing the source (observing windows). Pointing schemes are defined by specifying the prescription (observing mode) used to place the observing windows along the light curve, by the total amount of time available for observation ($T_{\text{obs}}$), and by the duration of each window ($T_{\text{win}}$). Two different observing modes are considered: consecutive and random observations. In the case of consecutive observations the source is observed once in every night, starting from the first night. For the random observing mode, windows are created at random positions inside randomly chosen nights. Although in this case there could be more than one window inside the same night, no overlap between the windows is permitted. In both cases, observing windows are constructed so that the summed duration of all windows is equal to the total observing time $T_{\text{obs}}$. If necessary the last window is resized in order not to exceed $T_{\text{obs}}$.

The main difference between the two observing modes is that while consecutive observations, especially when the observing windows are large, tend to be concentrated in a portion of the light curve, random observation are spread over the entire length of the simulated light curves. This translates into a different coverage of the Fourier spectra of the light curves, with the consecutive observing mode partially lacking coverage of the lowest frequencies. 

\subsection{Instrument response functions}

The instrument is modeled through the use of its instrument response functions (IRFs) for effective area and background rate. We make use of the IRFs that are publicly available for the CTA southern array~\cite{ctasite}. These IRFs are computed for a source at 20 degrees zenith angle, and have been optimized for observing times of 0.5, 5, and 50 hours.

Since the points of the simulated light curves represent the integral flux of the sources above $E_0=100$~GeV, the IRFs are integrated above this energy. In order to convert the points of the light curves into the rate of signal events above $E_0$, the differential effective area $A_{\text{eff}}(E)$ is folded with the differential flux of the simulated sources, here assumed to be a power-law with photon index $\Gamma=3$, which is a typical shape for AGN VHE spectra\footnote{Although it is known that source spectra are typically harder during flares, time variations of the spectral index are not considered in this work. The value $\Gamma=3$ is chosen as a compromise between softer spectra observed during quiescent states and the harder ones during flares.}. Defining $A_{\text{eff}}$ as 
\begin{equation}\label{eqn:atot}
A_{\text{eff}}=\frac{\displaystyle\sum_{E_i\geq E_0} A_{\text{eff}}(E_i)E_{i}^{-\Gamma}} {\displaystyle\sum_{E_i\geq E_0} E_i^{-\Gamma}},
\end{equation}
where $A_{\text{eff}}(E_i)$ is the effective area at energy $E_i$, the integral rate of signal events at a time $t_i$ can be written as $R_{\gamma}(t_i)=A_{\text{eff}}\phi(t_i)$. The expected rate of background events above $E_0$ is simply given by:
\begin{equation}\label{eqn:bgrate}
R_{\text{bg}}=\displaystyle\sum_{E_i\geq E_0} R_{\text{bg}}(E_i),
\end{equation}
with $R_{\text{bg}}(E_i)$ is the expected rate of background events at energy $E_{i}$. For the considered IRFs and photon index, the computed values of $A_{\text{eff}}$ and $R_{\text{bg}}$ are summarized in table~\ref{tab:irfs}.

\begin{table}[h]
\begin{center}
\begin{tabular}{c|c|c}
Optimization time {[}h{]}& $A_{\text{eff}}~{[}10^9~\mbox{cm}^2{]}$ & $R_{\text{bg}}~{[}\mbox{mHz}{]}$ \\
\midrule
0.5 & 1.97 & 50 \\
5 & 1.96 & 62 \\
50 & 1.50 & 37 \\
\end{tabular}
\caption{Integrated IRFs for the different optimization times.}
\label{tab:irfs}
\end{center}
\end{table}

These IRFs have been computed assuming a fixed zenith angle of 20 degrees, but they are a good approximation of the performance of the instrument in the range $[0, 30]$ degrees zenith. Having restricted the visibility windows to period where the source altitude is greater than 60 degrees ensure a consistent picture.

\section{Quantifying observing strategies performance}\label{sec:obs_quant}

Using the integrated IRFs described above and the simulated light curves we compute, for each observing strategy, the probability of detecting a source with given variability characteristics.

For both random and consecutive observing modes, $T_{\text{obs}}$ is varied between a minimum and a maximum observing time, $T_{\text{obs}}^{\text{min}}$ and $T_{\text{obs}}^{\text{max}}$ respectively and, for each value of $T_{\text{obs}}$, the window length takes values between $T_{\text{win}}^{\text{min}}$ and the minimum between $T_{\text{win}}^{\text{max}}$, and $T_{\text{obs}}^{\text{max}}$.

$T_{\text{obs}}^{\text{min}}$ and $T_{\text{obs}}^{\text{max}}$ are chosen in such a way as to focus the exploration of the parameter space in the region in which the probability of detection changes from almost zero to almost unity, the most interesting in the context of observing strategies optimization. We choose $T_{\text{obs}}^{\text{min}}=0.5$ hours, and $T_{\text{obs}}^{\text{max}}=10$ hours. The maximum allowed length for the observing windows is $T_{\text{win}}^{\text{max}}=4~\mbox{hours}$; the shortest windows will be $T_{\text{win}}^{\text{min}}=5~\mbox{min}$ long.

The ranges for the observing time and window length are covered using logarithmic bins. The bins of the simulation will have the same relative width: $\delta_{T_{\text{obs}}}$ and $\delta_{T_{\text{win}}}$. These determine the precision of the simulation: $\delta_{T_{\text{win}}}=20\%$, $\delta_{T_{\text{obs}}}=2\%$. Smaller values of $\delta_{T_{\text{win}}}$ can not be achieved, due to the 1-minute spacing of the points of the light curves.

Taking into account the time lost to re-point the instrument from one position in the sky to another, the overall time usage of an observing strategy is estimated as $T_{tot}=T_{\text{obs}}+N_{win}\Delta t_{slew}$ with $N_{win}$ being the number of windows employed and $\Delta t_{slew}$ the time it takes to re-point the telescopes, taken to be 1 minute for CTA medium-size telescopes\footnote{This is a rather conservative assumption, corresponding to the time it takes to point the telescopes to any direction in the sky. In a realistic scenario, observations are planned in such a way to minimize these losses.}.

For each observing strategy and set of light curves we apply the same sequence of observing windows to all the light curves in the set. Using the simulated fluxes and the integrated IRFs we compute the number of detected signal and background events as a function of the observing time along the light curve. At each time step $t_i$ we use the IRFs whose optimization time is closer to $t_i$ and compute the statistical significance of the gamma-ray excess in the signal region using Equation 17 from Li and Ma~\cite{liandma} taking $\alpha$, the ratio between signal and background region, to be $1/5$, as customary for IACTs analysis. If the significance is greater than $5$ and the number of excess signal events is greater than 10 for any point in the light curve, then the simulated source is flagged as detected.

For a given set of light curves and observing strategy, the number of light curves for which the source is detected, $N_{\text{det}}$, can be regarded as the number of successes obtained from a series of $N_{\text{LC}}=1000$ identical and independent yes/no tests; it therefore follows a binomial distribution with a probability of success $p_{\text{det}}=N_{\text{det}}/N_{\text{LC}}$, i.e. the probability of detecting the source with that particular strategy. We use the method proposed by Feldman and Cousins~\cite{feld_cou}, adapted to the binomial distribution, to construct confidence intervals that are always physically acceptable, i.e. that never violate the condition $0\leq p_{\text{det}} \leq1$. For each observing strategy, confidence intervals at the 3$\sigma$ confidence level have been computed.

The detection probability describes the performance of an observing strategy with respect to a given light curve type. For each observing strategy, the values of $p_{\text{det}}$ for the different types of variability are combined together to assess how that strategy behaves with respect to the variability properties of the sources. For a given observing strategy, let $p_{\text{det}}(F_{\text{rms}},\beta)$ be the probability of detecting sources whose variability characteristic are described by $F_{\text{rms}}$ and $\beta$. The number $N_{95\%}$ of combinations of $F_{\text{rms}}$ and $\beta$ for which $p_{\text{det}}\geq 95\%$ is given by:
\begin{equation}\label{eqn:noverth}
N_{95\%}=\sum_{F_{\text{rms}},\beta} H( p_{\text{det}}(F_{\text{rms}}, \beta) - 95\%),
\end{equation}
where $H(x)$ is the Heaviside step function and the sum runs through all of the considered values of $F_{\text{rms}}$ and $\beta$: ten values of $F_{\text{rms}}$ ($N_{F_{\text{rms}}}=10$), and eleven values of $\beta$ ($N_{\beta}=11$). How well an observing strategy performs over the whole variability parameter space is estimated with the parameter $R_{\text{cover}}$, defined as $R_{\text{cover}}=N_{95\%}/(N_{F_{\text{rms}}}N_{\beta})$. For a given observing strategy, $R_{\text{cover}}$ measures the fraction of the variability parameter space that is covered at high detection probability by that strategy. 
 
\section{Selection of optimal observing strategies}\label{sec:results}

Using the detection probability $p_{\text{det}}$, and $R_{\text{cover}}$, the performances of different observing strategies are compared and the ones more suited to accomplish a given task are identified. This method is applied here to search for optimal observing strategies in two different cases: detect a source with known variability properties in as little time as possible, and conduct variability-unbiased sky surveys. 

\subsection{Time-efficient observing strategies}

For a given light curve type, studying how $p_{\text{det}}$ varies as a function of $T_{\text{win}}$, $T_{\text{obs}}$, and observing mode, enables the identification of the fastest way to detect the source. The method described here can be used to tailor the observing strategy to individual sources whenever some indications of their variability properties are available, as for example in the case of pointed observations of previously detected AGN.

\begin{figure}[htbp]
	\begin{minipage}[b]{0.5\textwidth} 
	\includegraphics[width=\textwidth]{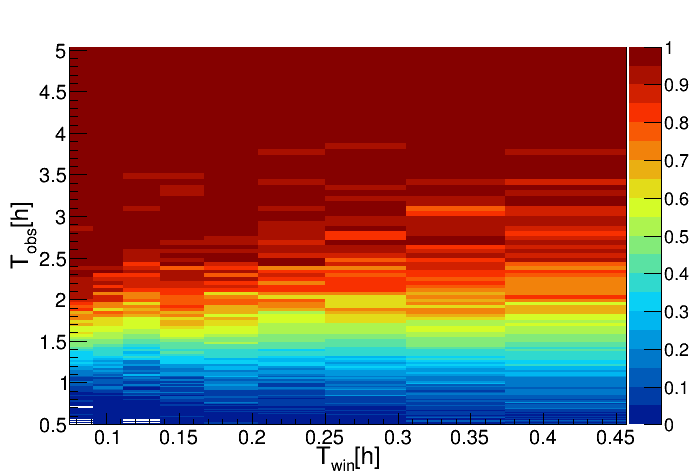}
	\end{minipage}
	\hfill
	\begin{minipage}[b]{0.5\textwidth}
	\includegraphics[width=\textwidth]{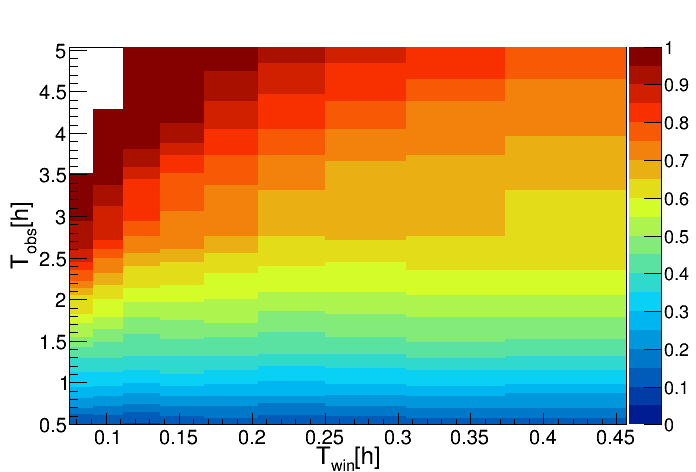}
	\end{minipage}
	\caption{$p_{\text{det}}(T_{\text{win}}, T_{\text{obs}})$ for random (left) and consecutive (right) observations for light curves characterized by $\beta=1.5$, $F_{\text{rms}}=50\%$. Note the changes of the $T_{\text{win}}$ axis range in the left panel. The empty region visible in the right panel is due to the fact that, for the consecutive observing mode, the visibility profile does not allow to reach large values of $T_{\text{obs}}$ when $T_{\text{win}}$ is small.}
	\label{fig:pdet_tobstwinall}
\end{figure}

To illustrate the main features of $p_{\text{det}}$ as a function of $T_{\text{win}}$, $T_{\text{obs}}$, and observing mode, we refer to medium variability light curves, characterized by $\beta=1.5$, and $F_{\text{rms}}=50\%$, see Figure~\ref{fig:pdet_tobstwinall}. Besides the obvious increase of $p_{\text{det}}$ with $T_{\text{obs}}$, the detection probability visibly depends on $T_{\text{win}}$. Dividing the observing time in many small windows increases the chances of observing the source during periods of high activity. Such observations have a high signal to noise ratio, increasing the overall significance and leading to a faster detection of the source. This effect is a consequence of the variability of the source, and becomes more marked as the source variability increase. For low-variability sources, the length of the observing windows has practically no influence on the detection probability, the differences between the two observing mode vanish, and $p_{\text{det}}(T_{\text{obs}})$ approaches the step function expected in the case of constant sources.

\begin{figure}[htbp]
	\begin{minipage}[b]{\textwidth} 
	\includegraphics[width=\textwidth]{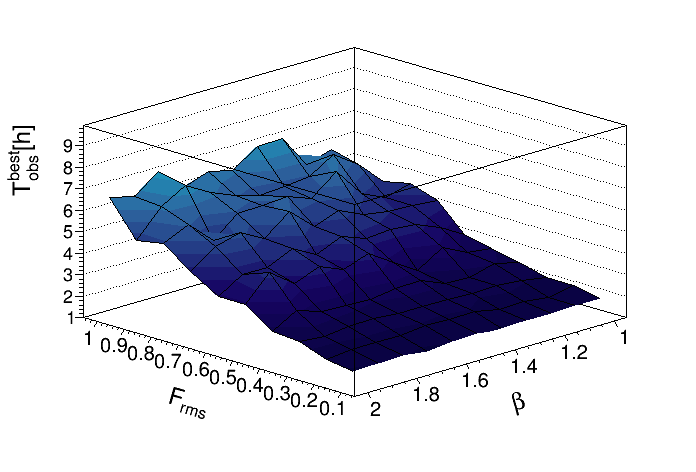}
	\end{minipage}
	\vfill
	\begin{minipage}[b]{\textwidth}
	\includegraphics[width=\textwidth]{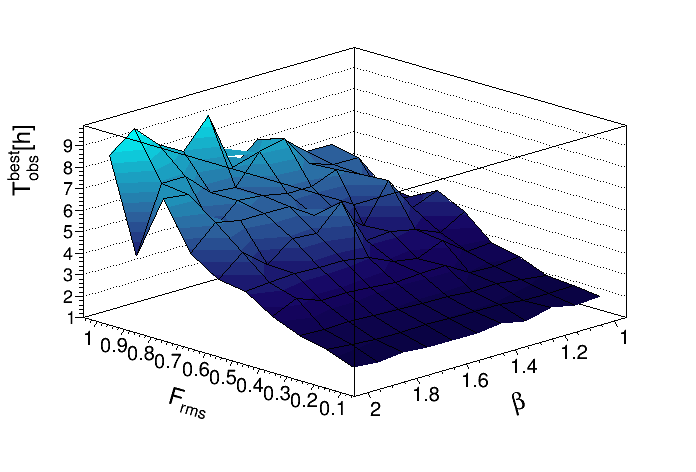}
	\end{minipage}
	\caption{The observing time for the best strategy, $T_{\text{obs}}^{best}[\mbox{h}]$, for random (top) and consecutive (bottom) observations, as a function of PSD index $\beta$ and fractional RMS amplitude $F_{\text{rms}}$ of the light curves.}
	\label{fig:best_tobstwinall}
\end{figure}

Time-efficient observing strategies are defined as the ones for which $p_{\text{det}}$ is compatible with one at 3-sigma confidence level, while having the smallest $T_{tot}$ and largest $T_{\text{win}}$ possible. With this definition, the parameters ($T_{\text{win}}^{best}$, $T_{\text{obs}}^{best}$) that characterize the best observing strategy are found for each light curve type and observing mode. The results are presented in Figure~\ref{fig:best_tobstwinall}, showing how $T_{\text{obs}}^{best}$ varies as a function of the parameters governing the variability: $F_{\text{rms}}$ and PSD index $\beta$. For both observing modes, $F_{\text{rms}}$ has the largest influence on the observing strategies. For low values of $F_{\text{rms}}$, $T_{\text{obs}}^{best}$ is of the order of $\sim1.5$~h, compatible with the time needed to detect a constant source with a flux equal to the mean flux of the simulated light curves. $T_{\text{obs}}^{best}$ increases with the variance of the light curves, i.e. sources with large amounts of total variability power are harder to detect. The PSD index has a weaker influence on the parameters of the best observing strategies. For the consecutive observing mode, the fastest way to detect a source almost exclusively uses the smallest observing windows considered, $T_{\text{win}}^{best}=5$~minutes. For random observations, only in the case of low fractional RMS ($F_{\text{rms}} \lesssim 30\%$), larger windows, of the order of $ \gtrsim 0.5$~h, can be used with success.

Random observations, as opposed to consecutive ones, are slightly faster in detecting the source, with a resulting time gain, averaged over the entire variability parameter space of $0.4$~h. However, for $F_{\text{rms}}\leq30$\%, the two observing modes have similar performances. Random observing windows are distributed along the entire length of the light curve, and variability components at all frequencies are sampled. Consecutive observations, especially when $T_{\text{win}}$ is large, are concentrated at the beginning of the light curve, thus partially lacking the coverage of the lowest frequencies. Since for $\beta>1$ most of the variability power and fluctuations are at low frequencies, for a fixed $T_{\text{obs}}$ consecutive observations, as compared to random ones, yield lower detection probability and are more influenced by the length of the windows.

\subsection{Observing strategies for sky surveys}

The parameter $R_{\text{cover}}$ is used to evaluate how much the performance of an observing strategy is influenced by the variability properties of the source. We use this information to select observing strategies that are less influenced by the variability properties of the sources, and are therefore more suited to conduct blind-sky surveys, where any observation bias, possibly induced by variability properties, has to be minimized. As we will see, the choice of a particular observing strategy can favor or disfavor the detection of the source, depending on its variability characteristics. Since there is evidence that different sub-classes of blazars have different variability behaviors~\cite{fermi_2lat_cat}, significant bias can be introduced in the sample of detected sources, if the survey strategy is not chosen properly.

\begin{figure}[htbp]
	\begin{minipage}[b]{0.5\textwidth} 
	\includegraphics[width=\textwidth]{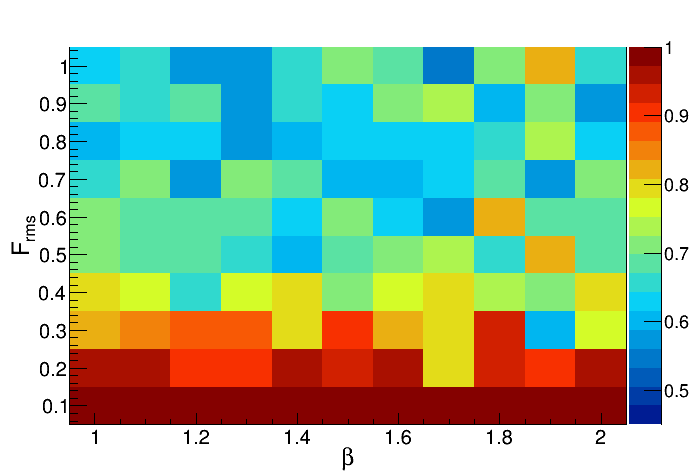}
	\end{minipage}
	\hfill
	\begin{minipage}[b]{0.5\textwidth}
	\includegraphics[width=\textwidth]{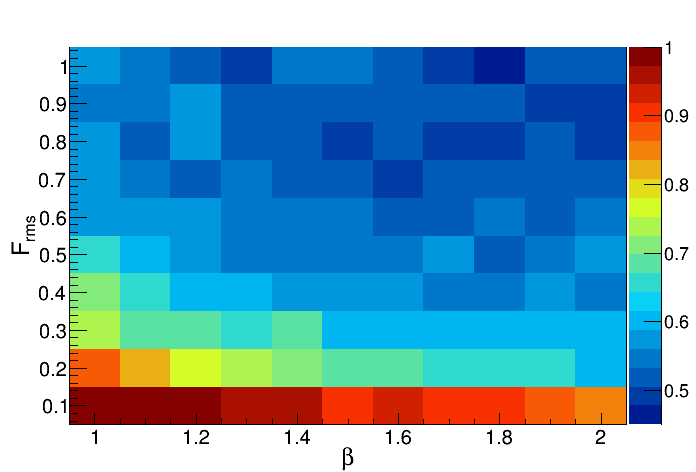}
	\end{minipage}
	\caption{$p_{\text{det}}$ vs. $F_{\text{rms}}$ and $\beta$ for observation with $T_{\text{obs}}=2$~h and $T_{\text{win}}=30$~min with random (left) and consecutive (right) observing mode. The corresponding values of $R_{\text{cover}}$ can be read directly from these plots, dividing the number of bins for which $p_{\text{det}}>95\%$ by the total number of bins (110).}
	\label{fig:biashisto_referenceobs}
\end{figure}

To illustrate this effect, we refer to a typical observation with observing windows of 30 minutes and 2 hours observing time, and see how the probability of detection changes for the different types of variability considered. In Figure~\ref{fig:biashisto_referenceobs}, $p_{\text{det}}$ is presented as a function of $F_{\text{rms}}$ and $\beta$ for this particular choice of $T_{\text{win}}$ and $T_{\text{obs}}$ for both random and consecutive observing modes. It is apparent that, if one use these strategies to carry out, for example, a blind survey of the sky, the resulting sample of detected sources will be biased towards low values of fractional RMS amplitude and harder PSD of the light curves. Taking random observations, for which $R_{\text{cover}}\sim14\%$, will prove more efficient in reducing this effect as compared to consecutive observations for which $R_{\text{cover}}\sim5\%$.

\begin{figure}[htbp]
	\begin{minipage}[b]{0.5\textwidth} 
	\includegraphics[width=\textwidth]{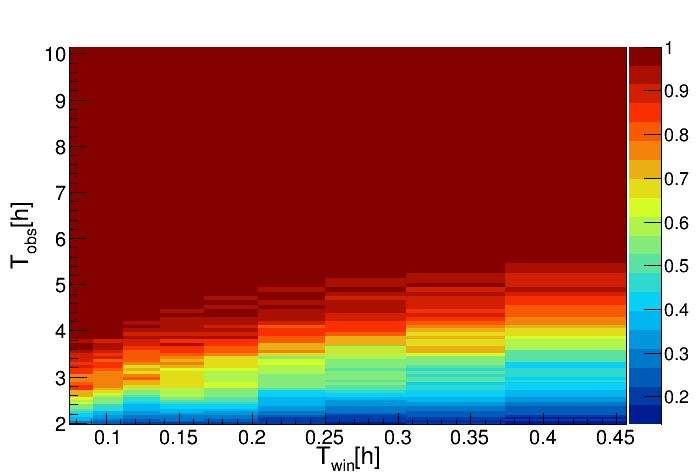}
	\end{minipage}
	\hfill
	\begin{minipage}[b]{0.5\textwidth}
	\includegraphics[width=\textwidth]{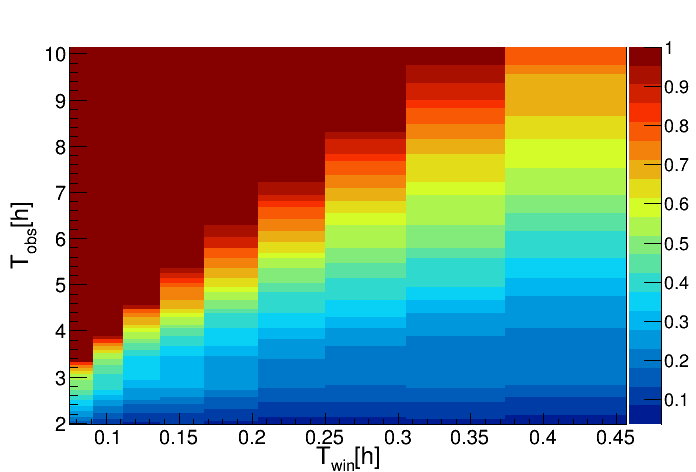}
	\end{minipage}
	\caption{$R_{\text{cover}}$ vs. $T_{\text{win}}$ and $T_{\text{obs}}$ for random (left) and consecutive (right) observing mode.}
	\label{fig:biashisto_alllc}
\end{figure}

The best observing strategy to conduct a survey is found asking for $R_{\text{cover}}=1$ and the smallest $T_{tot}$ possible. Such a strategy provides the fastest way to collect a sample of sources, in which the bias towards any kind of variability is lower than 5\%. Figure~\ref{fig:biashisto_alllc} presents the behavior of $R_{\text{cover}}$ in the observing strategies parameter space. As can be seen, for both observing modes, the most time-efficient way to conduct a variability-unbiased blind sky survey makes use of very small observing windows of $\sim5$~minutes. Using such small windows for each individual pointing of the survey, and a total time spent on each field of view of $\geq3.5$~hours\footnote{The study presented here is based on the on-source performace of the instrument. The time stated here has to be understood as on-source equivalent time.} there is little difference between random and consecutive observing modes.

\section{Summary and conclusions}

The performance of a time-limited pointing observatory for detection of sources characterized with AGN-like variability have been studied. Taking CTA as an example of such an instrument, observing strategies have been optimized for two scenarios: fast detection of sources with known variability characteristic, and variability-unbiased blind sky survey. We have used the current knowledge of AGN variability at high and very-high energy to simulate AGN-like light curves with different variability properties. Different observing strategies have been modeled and applied to these light curves and their performances are quantified through the probability of detection. Statistical errors associated with this probability are computed with the appropriate confidence belt construction.

We have shown that many short observations, spread over long periods of times, have several advantages over strategies that concentrate the observations in few long windows. In the former case the observations cover a wider range of variability time scales and have a higher probability of observing the sources during periods of high activity. For these reasons, such strategies are faster in detecting the sources. If small observing windows are used, the two observing modes that have been tested have similar performances. Even taking into account time losses in telescope slewing, for sources with substantial variability ($F_{\text{rms}}\simeq50\%$), observing strategies with $\sim$5-minute observing windows enables the savings of up to $\sim30$\% of the total observing time, as opposed to observing strategies with half an hour windows. Observing strategies using very small observing windows are also less influenced by the variability properties of the sources, which makes them more appropriate to conduct blind surveys with the least bias towards a certain source class. We have shown that an unbiased survey at 1\% of the Crab level can be achieved observing each field of view for a total of $\sim3.5$ hours with 5-minute windows. With the same observing time per field, but with half an hour observing windows, only low-variability ($F_{\text{rms}}\lesssim20\%$) sources will be detected.

\section*{Acknowledgments}
The authors thank the CTA Speaker And Publication Office for the courtesy review of this manuscript, and the CTA Consorium for the input provided. Thanks also to Nathan Kelley-Hoskins for the kind help and stimulating questions, to Rolf B\"{u}hler and Jonathan Biteau for the useful comments, and to Markus Paul Lindhout. 

\section*{References}
\thispagestyle{empty}
\bibliographystyle{ieeetr}
\bibliography{ggm_obsstrat}
\end{document}